\documentclass[journal,onecolumn]{IEEEtran}
\usepackage[cmex10]{amsmath}
\usepackage{amssymb}
\usepackage{amsfonts}
\usepackage{graphicx,color}
\usepackage{epsfig}
\usepackage{rotating}
\usepackage{multirow}
\usepackage{subfig}
\usepackage{cite}

\usepackage{caption}
\newtheorem{prop}{Proposition}

\begin{document}

\title{Joint Target Detection, Tracking and Classification with Forward-Backward PHD Smoothing}

\author{Yanyuan~Qin
}

\maketitle

\begin{abstract}
Forward-backward Probability Hypothesis Density (PHD) smoothing is an efficient way for target tracking in dense clutter environment. Although the target class has been widely viewed as useful information to enhance the target tracking, there is no existing work in literature which
incorporates the feature information into PHD smoothing. In this paper, we generalized the PHD smoothing by extending the “general mode”, which includes kinematic mode, class mode or their combinations etc., to forward-backward PHD filter. Through a “top-down” method, the general mode augmented forward-backward PHD smoothing is derived. The evaluation results show that our approach out-performs the state-of-art joint detection, tracking and classification algorithm in target state estimation, number estimation and classification. 
The reduction of OSPA distance is up to 40\%.
\end{abstract}

\begin{IEEEkeywords}
Joint target detection, tracking and classification, Probability Hypothesis Density (PHD), Forward-Backward Smoothing
\end{IEEEkeywords}

\IEEEpeerreviewmaketitle

\section{Introduction}
Multi-target tracking is a challenging problem for surveillance system. It is not only need estimate a set of dynamic target state but also targets numbers and types. Recently, probability hypothesis density (PHD) filter has received much attention~\cite{wei-a, vo-a,vo-b,vo-c,vo-d,jing2013current}. Compared with traditional association-based multi-target tracking approaches like joint probability data association (JPDA) and multiple hypothesis tracking (MHT), PHD filter avoids the data association between the measurement and track due to measurements and false alarms represented as random sets. The PHD filter is capable of dealing with the problem of target birth, spawning, disappearance in dense clutters.

Currently, there are two kinds of PHD implementations: sequential Monte Carlo PHD~\cite{vo-b} and Gaussian mixture PHD filter~\cite{vo-c}. Compared with Gaussian mixture PHD filter, the sequential Monte Carlo approximation is capable to handle the non-linear non-Gaussian problems. To track maneuvering target, multi-model sequential Monte Carlo PHD filter are introduced in .
Smoothing technique uses the observation in the future to improve the current state estimation precision. ~\cite{nandakumaran2007a}  derives the PHD smoothing based on finite set statistics method, respectively. It also can be extended to multi-model cases and the multi-model PHD smoothing is obtained. However, Mahler points out that existing works have adopted a bottom-up theoretical approach. That is, they take the PHD filter or the CPHD filter as starting point, and then attempt to generalize it to jump-Markov systems. None have adopted a theoretically top-down approach, which begin with the multi-target Bayes filter as the starting point; generalize it to a multi-target jump-Markov filter; and, then and only then, derive PHD filter equations from this generalized filter. As a result, it is unclear whether any of these proposed jump- Markov PHD filters are fully rigorous from a multi-target-statistics point of view. ~\cite{mahler2012a} derived a JMNS version of the PHD filter on multi-target jump-Markov systems through a top-down method.

The feature or class information of the target is useful to improve the performance of tracking~\cite{li2007a}. By incorporating the feature information to particle filter, JPDA and MHT, these methods are more efficient to track closely spaced parallel moving target or crossing moving targets from different classes. Random finite set (RFS) theory also provides an efficient tool to incorporate the feature information. If the feature measurements are considered, RFS-based PHD filter can also be applied to joint detection, tracking and classification~\cite{mahler-a,jing2018multi,qin2013multi,li2018multi}. Yang~\cite{wei-a} proposed to assign a class-matched PHD-like filter to each type of target, which has a class-dependent kinematic model set to describe the kinematic feature of targets precisely.

PHD smoothing improves the performance of PHD filter. However, to the best of our knowledge, the PHD smoothing with classification information has not been considered yet in literature. Therefore, we try to deal with this issue in this paper and derive the classification-aided PHD smoothing. Specifically, we derived the general feature conditioned forward-backward PHD smoothing through the “top-down” method.

In this paper, we make the following contributions:
\begin{itemize}
	\item  The forward-backward PHD-JDTC smoothing is proposed using a "top-down" approach, which can effectively deal with the multi-target joint detection, tracking and classification problem.
	\item The Sequencial Monte Carlo (SMC) implementation for forward-backward PHD-JDTC smoothing is presented and the stucture of PHD-JDTC smoothing is analyized.
	\item We propose to utilize the signal amplitude of the unknown SNR target for JDTC problem and avoiding the a priori information of the target average SNR. The simulation results show that our approach can make decision on class information effectively and outperform the state-of-art approach --- PHD-JDTC filter in target tracking accuracy.
\end{itemize}

This paper is organized as follows. First, Section~\ref{sec:PHDsmoothing} provides a brief review of the forward-backward PHD smoothing based on Random Finite Set theory. It also presents the forward PHD filter with class-dependent kinematic model set. In Section~\ref{sec:PHDgeneralmode}, PHD filter and smoothing with general feature are derived from a “top-down” approach. In Section ~\ref{sec:phdjdtc}, the forward-backward PHD-JDTC smoothing is derived by considering the class information. In addition, its Sequential Monte Carlo implementation is provided. In Section~\ref{sec:evaluation}, simulation case is designed and the evaluation results shows the performance of proposed approach. Finally, conclusion and future work are given in Section~\ref{sec:conclusion}.

\section{RFS-based forward-backward PHD smoothing}\label{sec:PHDsmoothing}
In multi-target tracking, both the target number and states are random, as well as the number of measurements and the measurements themselves. Therefore the states and measurements could be model by Random Finite Set. This section reviews the forward-backward PHD smoothing, which provides the basis for deriving the forward-backward PHD smoothing with augmented general mode. In Subsection~\ref{subsec:bayes}, the Random Finite Set are used to model the multi-target states and observations. The multi-target Bayes filter and smoothing are given. Its first-order approximate, PHD smoothing algorithm is outlined in Subsection~\ref{phdsmoothing}. 

\subsection{Multi-target Bayes forward-backward smoothing}\label{subsec:bayes}

Assume that at time $k$, single target state $x_k$ belongs to the state space $E_s$, i.e. $x_k \in E_s $, then the multi-target states can be defined as follows:
\begin{equation}{X_k} \buildrel \Delta \over = \left\{ {{x_{k,1}},{x_{k,2}}, \cdots {x_{k,{N_k}}}} \right\} \in {\cal F}\left( {{E_s}} \right)\end{equation}
Suppose that the single target observation space is $E_o$, single target observation at time $k$ is $z_k \in E_o$ , the multi-states is given by
\begin{equation}{Z_k} \buildrel \Delta \over = \left\{ {{z_{k,1}},{z_{k,2}}, \cdots {z_{k,{M_k}}}} \right\} \in {\cal F}\left( {{E_o}} \right)\end{equation}
where $M_k$ and $N_k$ are the number of targets and their measurements, respectively, while ${\cal F}({E_s})$and ${\cal F}({E_o})$ are the finite subsets of $E_s$ and $E_o$, respectively. 

For the measurements from time 1 to $k$, the aggregate of all of them can be expressed as

\begin{equation}{{\rm Z}_{1:k}} = \bigcup\limits_{i = 1}^k {{Z_i}} \end{equation}
In the framework of FISST (Finite Set Statistics), the uncertainty for multi-target states and the corresponding observations are represented by random finite sets.

Forward-backward smoothing consists of forward filtering followed by backward smoothing. In the forward filtering, the posterior density is propagated forward to time k via Bayes recursion. In the backward smoothing step, the smoothed density is propagated backward, from time k<k, via the backward smoothing recursion. Analogy to the single target Bayes filter predictor, multi-target forward predication is calculated as follows
Forward-backward smoothing consists of forward filtering followed by backward smoothing. In the forward filtering, the posterior density is propagated forward to time $k$ via Bayes recursion. In the backward smoothing step, the smoothed density is propagated backward, from time k<k, via the backward smoothing recursion. Analogy to the single target Bayes filter predictor, multi-target forward predication is calculate as follows,

\begin{equation}
{D_{k|k - 1}}({X_k}|{{\rm Z}_{1:k - 1}}) = \int {{f_{k|k - 1}}({X_k}|X){D_{k - 1|k - 1}}(X|{{\rm Z}_{1:k - 1}})\delta } X
\end{equation}

where $\int$ and $\delta$ represents random set integral and differential.  ${f_{k|k - 1}}({X_k}|X)$ is the multi-target Makov density. 

With the measurements from time $k$, multi-target forward update is given by, 
\begin{equation}
D_{k|k}(X_k| Z_{1:k}) =
\frac{{{h_k}({Z_k}|{X_k}){D_{k|k - 1}}({X_k}|{{ Z}_{1:k - 1}})}}
{\int {{h_k}({Z_k}|X){D_{k + k - 1}}(X|{ Z}_{1:k - 1})\delta X} }
\end{equation}

where ${h_k}({Z_k}|{X_k})$ is the multisource likelihood function.

The smoothed multi-target density is propagated backward, from time $k$ to $k' < k$, via the multi-target backward smoothing recursion.
\begin{equation}
{D_{t|k}}({X_t}|{Z_{1:k}}) = {D_{t|t}}({X_t}|{Z_{1:t}})\int {{f_{t + 1|t}}({X_{t + 1}}|{X_t})\frac{{{D_{t + 1|k}}({X_{t + 1}}|{Z_{1:k}})}}{{{D_{t + 1|t}}({X_{t + 1}}|{Z_{1:t}})}}\delta {X_{t + 1}}} 
\end{equation}

\subsection{PHD filter and smoothing}\label{phdsmoothing}
For multi-target Bayes filter, it is intractable to implement in a computational manner. Under the assumption that no target generates more than one measurement and each measurement is generated by no more than a single target, all measurements are conditionally independent of target state, missed detections, and a multi-object Poisson false alarm process, Mahler~\cite{mahler-a} proposed first-order multi-target moment approximation for multi-target Bayes filter- Probability Hypothesis Density (PHD). Given any region $S$ of single-target state space $X_0$, the integral 
$\int_S {{D_{k|k}}(x)dx} $ is the expected number of targets in $S$. In particular, if $S = X_0$ is the entire state space then ${N_{k|k}} \buildrel \Delta \over = \int {{D_{k|k}}(x)dx} $ is the total expected number of targets in the scene.

Compared with optimal multi-target tracking Bayes recursion, PHD filter is much easier because of its first-order multi-target moment approximation. On the other hand, the computational complexity is small as the integral of PHD filter is performed on the single target space. 

PHD forward-backward smoothing could be derived from physical-space approach~\cite{nadarajah-a} and standard point process theory~\cite{vo-d}, respectively. 

\textbf{PHD forward filtering}.
In prediction step,
\begin{equation}
{D_{k + 1|k}}(x) = {\gamma _{k + 1|k}}(x) + \int {{F_{k + 1|k}}(x|x') \cdot {D_{k|k}}\left( {x'} \right)} dx'
\end{equation}
where 
${F_{k + 1|k}}(x|x') = {P_s}(x'){f_{k + 1|k}}(x|x') + {\beta _{k + 1|k}}(x|x')$, 
$\gamma_k (x)$ is the intensity of birth targets at time $k$, ${\beta _{k|k - 1}}(x|x')$ is the intensity of spawned targets. 
${P_s}(x')$ is the survive probability for existing targets. 

The state is updatd then 
\begin{equation}
{D_{k + 1|k + 1}}(x) = {D_{k + 1|k}}(x)\left[ {1 - {P_D}(x) + {P_D}(x)\sum\limits_{z \in Z} {\frac{{{L_z}(x)}}{{{\kappa _k}(z) + {D_{k + 1|k}}\left[ {{P_D}(x){L_z}(x)} \right]}}} } \right],
\end{equation}
where $P_D(x)$ is the probability of detection for targets and $\kappa_k(z)$  is the intensity for clutters. 

\textbf{PHD backward smoothing}. Smoothing provides more improved estimation results than filtering due to that it make use of more measurements. There are mainly three kinds of smoothing techniques: fixed-interval smoothing, fixed-point smoothing and fixed-lag smoothing. Fixed lag smoothing estimate the state at time  given measurements until for fixed-time lag. Here we consider the fixed lag smoothing for multi-target Bayes backward recursion.

\begin{equation}
{D_{t|k}}({x_t}|{Z_{1:k}}) = {D_{t|t}}({x_t}|{Z_{1:t}})\left[ {{P_s}({x_t})\int {\frac{{{D_{t + 1|k}}({x_{t + 1}}|{Z_{1:k}}){f_{t + 1|t}}({x_{t + 1}}|{x_t})}}{{{D_{t + 1|t}}({x_{t + 1}}|{Z_{1:t}})}}d{x_{t + 1}} + 1 - {P_s}({x_t})} } \right],
\label{phdbackwardsmoothing}
\end{equation}
where 

${D_{t + 1|t}}({x_{t + 1}}|{Z_{1:t}}) = \gamma ({x_{t + 1}}) + \int {\left[ {{P_s}({x_t}){f_{t + 1|t}}({x_{t + 1}}|{x_t}) + {\beta _{t + 1|t}}({x_{t + 1}}|{x_t})} \right]} {D_{t|t}}({x_t})d{x_t}$.
It should be noted that the backward recursion is initialized with the filtering results at the present time $k$  and stopped at time $k-L$, where $L$ is the time lag of the smoothing algorithm.

\section{forward-backward PHD filter with general mode}\label{sec:PHDgeneralmode}
The PHD smoothing algorithm in section~\ref{sec:PHDsmoothing} only take into account the kinematic state of targets, which is not able to handle the issue of maneuvering target tracking or joint detection, tracking and classification. To tackle these problems, it is necessary to append the general mode, which might be the kinematic mode, the classification or the combination of them, to the target states. The general mode can be viewed as general jump Markov state, which is just related with the state at last time step while be independent with states of earlier time steps. In this section, the forward-backward PHD smoothing whose states extended with general jump Markov mode are given in details. First the forward extended PHD filter will be outlined in subsection~\ref{PHDfiltergeneralmode} , with the backward extended PHD smoothing recursion derived in subsection~\ref{PHDsmoothgeneral}. 

\subsection{Forward PHD filter with general mode}\label{PHDfiltergeneralmode}
\begin{prop}
 Kinematic mode and class information could be sort as general jump Markov mode.
\end{prop}

The single target state space consists of augmented states of the form $\ddot x = (x,o)$ , based on which the multi-target state have the form of 
$\ddot X = \left\{ {({x_1},{o_1}),...,({x_n},{o_n})} \right\}$, where $o$ is the general mode with jump Markov feature. Substitute  $x$ with $x_ddot$ for the PHD prediction and update equation, then we get the following expressions.

\begin{equation}
{D_{k + 1|k}}(\ddot x) = {\gamma _{k + 1|k}}(\ddot x) + \int {{F_{k + 1|k}}(\ddot x|\ddot x'){D_{k|k}}\left( {\ddot x'} \right)} d\ddot x'
\end{equation}
where 
${F_{k + 1|k}}(\ddot x|\ddot x') = {P_s}(\ddot x'){f_{k + 1|k}}(\ddot x|\ddot x') + {\beta _{k + 1|k}}(\ddot x|\ddot x')$.
The prediction step is 
\begin{equation}
{D_{k + 1|k + 1}}(\ddot x) = {D_{k + 1|k}}(\ddot x)\left[ {1 - {P_D}(\ddot x) + {P_D}(\ddot x)\sum\limits_{z \in {Z_{k + 1}}} {\frac{{{L_z}(\ddot x)}}{{{\kappa _{k + 1}}(z) + {\tau _{k + 1}}(z)}}} } \right],
\end{equation}
where ${\tau _{k + 1}}(z) = \int {{P_D}(\ddot x){L_z}(\ddot x){D_{k + 1|k}}(\ddot x)} d\ddot x$.

The corresponding expanded full jump-variable notation will be:
\begin{equation}
{D_{k + 1|k}}(x,o) = {\gamma _{k + 1|k}}(x,o) + \sum\limits_{o'} {\int {{F_{k + 1|k}}(x,o|x',o'){D_{k|k}}\left( {x',o'} \right)} dx'}. 
\end{equation}
where 
\begin{equation}
{F_{k + 1|k}}(x,o|x',o') = {P_s}(x',o'){f_{k + 1|k}}(x,o|x',o') + {\beta _{k + 1|k}}(x,o|x',o')
\end{equation}

\begin{equation}
{D_{k + 1|k + 1}}(x,o) = {D_{k + 1|k}}(x,o)\left[ {1 - {P_D}(x,o) + {P_D}(x,o)\sum\limits_{z \in {Z_{k + 1}}} {\frac{{{L_z}(x,o)}}{{{\kappa _{k + 1}}(z) + {\tau _{k + 1}}(z)}}} } \right]
\end{equation}

\begin{equation}
{\tau _{k + 1}}(z) = \sum\limits_o {\int {{P_D}(x,o){L_z}(x,o){D_{k + 1|k}}(x,o)} dx}.
\end{equation}
 Integrate out mode  and the PHD for target we can get multi-target state estimation,
\begin{equation}\begin{array}{l}
{D_{k + 1|k + 1}}(x) = \sum\limits_o {{D_{k + 1|k + 1}}(x,o)} 
= 1 - \sum\limits_o {{P_D}(x,o){D_{k + 1|k}}(x,o)}  + \sum\limits_{z \in {Z_{k + 1}}} {\frac{{\sum\limits_o {{P_D}(x,o){L_z}(x,o){D_{k + 1|k}}(x,o)} }}{{{\kappa _{k + 1}}(z) + {\tau _{k + 1}}(z)}}} 
\end{array}\end{equation}
And the expected target number is 
${N_{k + 1|k + 1}}(x) = \int {{D_{k + 1|k + 1}}(x)dx} $.

\subsection{Backward PHD filter with general mode}\label{PHDsmoothgeneral}
\begin{prop}
General mode extended PHD smoothing can be achieved.
\end{prop}
 After we get the estimation of target state at time $k$ we can use the do  backward smoothing by extending $o$ to $\ddot x = (x,o)$ 
\begin{equation}
{D_{t|k}}({x_t},o|{Z_{1:k}}) = {D_{t|t}}({x_t},o|{Z_{1:t}})
\times \left[ {{P_s}({x_t},o)\sum\limits_{o'} {\int {\frac{{{D_{t + 1|k}}({x_{t + 1}},o'|{Z_{1:k}}){f_{t + 1|t}}({x_{t + 1}},o'|{x_t},o)}}{{{D_{t + 1|t}}({x_{t + 1}},o'|{Z_{1:t}})}}d{x_{t + 1}} + 1 - {P_s}({x_t},o)} } } \right]
\end{equation}

Where 
\begin{equation}
{D_{t + 1|t}}({x_{t + 1}},o'|{Z_{1:t}}) = \gamma ({x_{t + 1}},o')
 + \sum\limits_o {\int {\left[ {{P_s}({x_t},o){f_{t + 1|t}}({x_{t + 1}},o'|{x_t},o) + {\beta _{t + 1|t}}({x_{t + 1}},o'|{x_t},o)} \right]} {D_{t|t}}({x_t},o)d{x_t}} 
\end{equation}

\textbf{Proof}: Substitute $x$ with $\ddot{x}$  in Eq.~\ref{phdbackwardsmoothing}, we get 
\begin{equation}
{D_{t|k}}({\ddot x_t}) = {D_{t|t}}({\ddot x_t})\left[ {{P_s}({{\ddot x}_t})\int {\frac{{{D_{t + 1|k}}({{\ddot x}_{t + 1}}){f_{t + 1|t}}({{\ddot x}_{t + 1}}|{{\ddot x}_t})}}{{{D_{t + 1|t}}({{\ddot x}_{t + 1}})}}d{{\ddot x}_{t + 1}} + 1 - {P_s}({{\ddot x}_t})} } \right]
\end{equation}

where 
\begin{equation}
{D_{t + 1|t}}({\ddot x_{t + 1}}) = \gamma ({\ddot x_{t + 1}}) + \int {\left[ {{P_s}({{\ddot x}_t}){f_{t + 1|t}}({{\ddot x}_{t + 1}}|{{\ddot x}_t}) + {\beta _{t + 1|t}}({{\ddot x}_{t + 1}}|{{\ddot x}_t})} \right]} {D_{t|t}}({\ddot x_t})d{\ddot x_t}
\end{equation}
Then, substitute  $\ddot x = (x,o)$  to the above equation.

Note that if we use $r$ to replace $o$, we will get the multi-model PHD filter and multi-model PHD smoothing, which is consistant with the work in~\cite{nadarajah-a}. 

\section{forward-backward PHD filter with class information}\label{sec:phdjdtc}
In this section, we will first introduce the PHD-JDTC filter and smoothing and then give the SMC implementation for it.
\subsection{PHD-JDTC Filter}
The jump Markov PHD filter is mainly designed for the case where the motion pattern of target change. In fact, however, the class information can also be viewed as a jump Markov variable. Specifically, the existing target is not changing with time, which is a special type of jump Markov variable. However, for the birth target, it might be different the parent target.  For instance, a missile is launched from aircraft. In~\cite{wei-a}, the class conditional PHD-JDTC filter is derived based on the addition of point process instensity function. In this section, we will show that PHD-JDTC filter can be easily derived by  the "top-down" approach based on PHD filter with general mode.

\begin{prop}
	Through the general mode extended method, the joint detection, tracking and classification PHD algorithm will be achieved via a “top-down” way. 
\end{prop}

Let's treat the class as a special kind of “mode” which is modeled as Jump-Markov model. If we augment the state $x$ with class information $c$, the class-conditioned PHD filter can be reached by substituting $x$ by  $\ddot{x} = (x,c)$ within Equation~\ref{phdbackwardsmoothing}. 

\begin{equation}
{D_{k + 1|k}}(x,c) = {\gamma _{k + 1|k}}(x,c) + \sum\limits_{c'} {\int {{F_{k + 1|k}}(x,c|x',c'){D_{k|k}}\left( {x',c'} \right)} dx'} \end{equation}

\begin{equation}
{D_{k + 1|k + 1}}(x,c) = {D_{k + 1|k}}(x,c)\left[ {1 - {P_D}(x,c) + {P_D}(x,c)\sum\limits_{\tilde z \in {{\tilde Z}_{k + 1}}} {\frac{{{L_z}(x,c)}}{{{\kappa _{k + 1}}(\tilde z) + {\tau _{k + 1}}(\tilde z)}}} } \right]
\end{equation}
where
\begin{equation}
{F_{k + 1|k}}(x,c|x',c') = {P_s}(x',c'){f_{k + 1|k}}(x,c|x',c') + {\beta _{k + 1|k}}(x,c|x',c')
 \end{equation}
\begin{equation}
{\tau _{k + 1}}(\tilde z) = \sum\limits_c {\int {{P_D}(x,c){L_{\tilde z}}(x,c){D_{k + 1|k}}(x,c)} dx}
 \end{equation}
Assume that the class of existing target does not change with time, i.e. 
\begin{equation}
{f_{k + 1|k}}(c|c') = \left\{ {\begin{array}{*{20}{c}}
{0,~~if ~~c \ne c'}\\
{1, ~~if ~~c = c'}
\end{array}} \right.
 \end{equation}
However, for the type of  spawning target, it might be a time-variant jump Markov state.

The equation now is in the following form:
In the state prediction step,
\begin{equation}\begin{array}{l}
{D_{k + 1|k}}(x,c) = {\gamma _{k + 1|k}}(x,c) + \sum\limits_{c'} {\int {{\beta _{k + 1|k}}(x,c|x',c'){D_{k|k}}\left( {x',c'} \right)dx'} } 
  + \int {{P_s}(x',c){f_{k + 1|k}}(x,c|x',c)dx'} 
  \label{equ:targetclass}
\end{array}
\end{equation}

Observation is updated as follows,
\begin{equation}\begin{array}{l}
{D_{k + 1|k + 1}}(x,c) = {D_{k + 1|k}}(x,c)
   \times {\kern 1pt} \left[ {1 - {P_D}(x,c) + {P_D}(x,c)\sum\limits_{\tilde z \in {{\tilde Z}_{k + 1}}} {\frac{{{L_{\tilde z}}(x,c)}}{{{\kappa _{k + 1}}(\tilde z) + \sum\limits_c {\int {{P_D}(x,c){L_{\tilde z}}(x,c){D_{k + 1|k}}(x,c)} dx} }}} } \right]
\end{array}
\end{equation}

The number of target can be calculated by summarize all the classes, i.e.,
\begin{equation}
{D_{k + 1|k + 1}}(x) = \sum\limits_c {{D_{k + 1|k + 1}}(x,c)} 
\end{equation}
Finally, we get the formulation of PHD-JDTC filter.

\subsection{PHD-JDTC smoothing}
\begin{prop}
The class-conditioned backward PHD filter is,
\begin{eqnarray}
{D_{t|k}}({x_t},c|{Z_{1:k}}) = {D_{t|t}}({x_t},c|{Z_{1:t}})\left[ {{P_s}({x_t},c)\sum\limits_{c'} {\int {\frac{{{D_{t + 1|k}}({x_{t + 1}},c'|{Z_{1:k}}){f_{t + 1|t}}({x_{t + 1}},c'|{x_t},c)}}{{{D_{t + 1|t}}({x_{t + 1}},c'|{Z_{1:t}})}}d{x_{t + 1}} + 1 - {P_s}({x_t},c)} } } \right],
\end{eqnarray}

where
\begin{equation}\begin{array}{l}
{D_{t + 1|t}}({x_{t + 1}},c' = c|{Z_{1:t}}) = \gamma ({x_{t + 1}},c') + \int {{P_s}({x_t},c){f_{t + 1|t}}({x_{t + 1}}|{x_t},c){D_{t|t}}({x_t},c|{Z_{1:t}})d{x_t}} \\
 {\kern 1pt} {\kern 1pt}  + \sum\limits_{c''} {\int {{\beta _{t + 1|t}}({x_{t + 1}},c' = c|{x_t},c''){D_{t|t}}({x_t},c''|{Z_{1:t}})d{x_t}} } .
\end{array}
\end{equation}
\end{prop}
\textbf{Proof}: we extend $x$ by adding the class information $c$,  i.e. $\ddot{x} =(x, c)$ and substuting to Equation~\ref{phdbackwardsmoothing}. We get  
\begin{equation}
{D_{t|k}}({x_t},c|{Z_{1:k}}) = {D_{t|t}}({x_t},c|{Z_{1:t}})
  \times \left[ {{P_s}({x_t},c)\sum\limits_{c'} {\int {\frac{{{D_{t + 1|k}}({x_{t + 1}},c'|{Z_{1:k}}){f_{t + 1|t}}({x_{t + 1}},c'|{x_t},c)}}{{{D_{t + 1|t}}({x_{t + 1}},c'|{Z_{1:t}})}}d{x_{t + 1}} + 1 - {P_s}({x_t},c)} } } \right]
\end{equation}

where 
\begin{equation}
\begin{array}{l}
{D_{t + 1|t}}({x_{t + 1}},c'|{Z_{1:t}}) = \gamma ({x_{t + 1}},c')
 + \sum\limits_c {\int {\left[ {{P_s}({x_t},c){f_{t + 1|t}}({x_{t + 1}},c'|{x_t},c) + {\beta _{t + 1|t}}({x_{t + 1}},c'|{x_t},c)} \right]} {D_{t|t}}({x_t},c)d{x_t}} 
\end{array}
\end{equation}
 Assume that the type of existing target does not change with time.  Using Equation ~\ref{equ:targetclass} we can get the PHD-JDTC smoothing.
 
 The final state estimation with smoothing will be
\begin{equation}
{D_{t|k}}({x_t}|{Z_{1:k}}) = \sum\limits_c {{D_{t|k}}({x_t},c|{Z_{1:k}})}
 \end{equation}

The number of target after smooth is 
\begin{equation}
{N_{t|k}}({x_t}) = \int {{D_{t|k}}({x_t}|{Z_{1:k}})d{x_t}} 
\end{equation}

\textbf{Remark}: In the stage of smoothing, different kind of targets could be processed in the class-conditioned PHD smoothings. On the other hand, it can also be seen that there is  information interaction in the proposed forward-backward PHD smoothing algorithm. In the forward filtering step, the spawning targets has mutual information exchange in the prediction phase, while different target interact with others by the joint likelihood function in update stage. It also has the information exchange in the backward smoothing due to that the spawning targets might be different with the target they spawned from. 

\subsection{SMC implementation}
\begin{figure}[!t]
	\centering
	\includegraphics[width=4.5in]{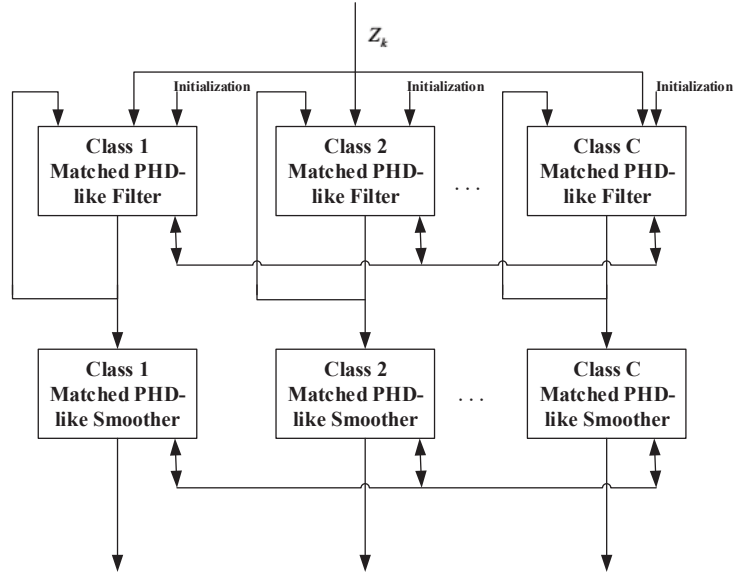}
	\caption{Recursive forward-backward PHD JDTC smoothing }
	\label{fig:JDTC}
\end{figure}

Based on the equations above, the forward-backward smoothing can be expressed with an explicit structure, from Fig.~\ref{fig:JDTC} we can see the  mutual information exchange between targets from different classes.

Next, we will present how to implement PHD-JDTC smoothing with SMC.  
The filtering step can be divided into three steps: particle prediction,  update and  resampling.

Assume the state vector of particle with class and model information is 
\begin{equation}
\dddot{x} \in \bigcup {_{c \in C}} ({X_c} \times c \times {M_c})
\end{equation}

\textbf{Step 1}: assume at time $k-1$, the PHD is $D_{k-1,c}$. The targets of class $c$ can be represented by $N_{k-1}^c$ equal weighted particles,
$\{ w_{k - 1,c}^s,\ddot x_{k - 1,c}^s\} _{s = 1}^{N_{k - 1}^c}$, the state of targets in different classes will update based on the state transfer model. 

For the existing targets, the particles are directly update their state within their own PHD filter, then the predicted class for those particles is
\begin{equation}c_{\beta ,k|k - 1}^s \sim \phi ( \cdot |{c_{k - 1}}),s = 1,...,{J_{\beta ,{c_{k - 1}}}}\end{equation}
where $c_{k-1}$ is the class for the original targets. $J_{\beta ,c_{k - 1}}$ is number of particles for spawn targets. Usually, the recommended distribution is set as same as the transfer probability of spawn targets $p_{\beta,k|k-1}(c|c')$. The weight of spawn target is
\begin{equation}w_{\beta ,k|k - 1}^s(c_{\beta ,k|k - 1}^s) = \frac{{{p_{\beta ,k|k - 1}}(c_{\beta ,k|k - 1}^s|c_{k - 1}^s)}}{{\phi (c_{\beta ,k|k - 1}^s|c_{k - 1}^s)}},s = 1,...,{J_{{c_{k - 1}}}}\end{equation}
For the manuver target, the prediction of model $r_{\beta ,k|k - 1}^s(c_{\beta ,k|k - 1}^s)$ will use the same random sampling approach. The predicted particles
$\left\{ {{x_{c_{\beta ,k|k - 1}^s}}} \right\}_{s = 1}^{{J_{{c_{k - 1}}}}}$ are sampled in  distribution ${\varphi ({x_{c_{\beta ,k|k - 1}^s}}|c_{\beta ,k|k - 1}^s,{x_{c_{k - 1}^s}},c_{k - 1}^s)}$ .  The weight of particles then becomes
\begin{equation}\tilde w_{\beta ,k|k - 1}^s(c_{\beta ,k|k - 1}^s) = w_{\beta ,k|k - 1}^s(c_{\beta ,k|k - 1}^s)\frac{{{\beta _{k|k - 1}}({x_{c_{\beta ,k|k - 1}^s}}|{x_{c_{k - 1}^s}})}}{{\varphi ({x_{c_{\beta ,k|k - 1}^s}}|c_{\beta ,k|k - 1}^s,{x_{c_{k - 1}^s}},c_{k - 1}^s)}},s = 1,...,{J_{{c_{k - 1}}}}\end{equation}
After the prediction step for all classes of particles, the particles for each class $c$ is represented as 
$\left\{ {\tilde w_{k|k - 1}^s,\dddot{x}_{k|k - 1,c}^s} \right\}_{s = 1}^{{{\tilde N}_{k|k - 1,c}}}$.

\textbf{Step 2}: Particle state upddate 

\begin{equation}\hat w_{k,c}^s = \tilde w_{k|k - 1,c}^s\left[ {(1 - {p_{D,c}}(\dddot{x}_{k|k - 1,c}^s)) + \sum\limits_{j = 1}^{M(k)} {\frac{{{p_{D,c}}(\dddot{x}_{k|k - 1,c}^s){g_{k,c,r}}(z_{k,\dddot{x}}^j|\dddot{x}_{k|k - 1,c}^s){h_{k,c,r}}(z_{k,f}^j|\dddot{x}_{k|k - 1,c}^s)}}{{K(z_{k,x}^j){c_f}(z_{k,f}^j) + \Psi (z_k^j)}}} } \right]
\end{equation}
where 
\begin{equation}\Psi (z_k^j) = \sum\limits_{c \in C} {\sum\limits_{s = 1}^{\tilde N_{k|k - 1}^c} {{p_{D,c}}(\dddot{x}_{k|k - 1,c}^s){g_{k,c,r}}(z_{k,x}^j|\dddot{x}_{k|k - 1,c}^s){h_{k,c,r}}(z_{k,f}^j|\dddot{x}_{k|k - 1,c}^s)} } \tilde w_{k|k - 1,c}^s
\end{equation}
${g_{k,c,r}}(z_{k,x}^j|\dddot{x}_{k|k - 1,c}^s)$ is the state measurement likelyhood function, and ${h_{k,c,r}}(z_{k,f}^j|\dddot{x}_{k|k - 1,c}^s)$ is the class measurement likelyhood function.

\textbf{Step 3}:Particle resampling 
After the state update, the corresponding particles of each category are re-sampled within the category to avoid particle depletion. Finally we get the updated particle collection
$\{ \bar w_{k,c}^s,\dddot{x}_{k,c}^s\} _{s = 1}^{N_k^c}$

\textbf{Step 4}:  Particle smoothing. 
We denote the Probability Hypophisis Density 
${D_{t|t}}({x_t},{r_t},{c_t}|{Z_{1:t}})$ as follows,
\begin{equation}
\{ w_{t|t}^{(s)},x_{t|t}^{(s)},r_{t|t}^{(s)},c_{t|t}^{(s)}\} _{s = 1}^{{L_t}},t = k - L,...,k
\end{equation}
For target that belongs to class $c$, we have PHD as
\begin{equation}
{D_{t|t}}({x_t},{r_t},c|{Z_{1:t}}) = \sum\limits_{s = 1}^{{L_t}(c)} {w_{c,t|t}^{(s)}\delta ({x_t} - x_{t|t}^{(s)},{r_t} - r_{t|t}^{(s)},c - {c^{(s)}})}
 \end{equation}
For steps $t = k-1,..., k-L$
\begin{equation}
w_{c,t|k}^{(s)} = w_{c,t|t}^{(s)}\left[ {{P_s}(x_t^{(s)})\sum\limits_{q = 1}^{{L_{t + 1}}(c)} {\frac{{w_{c,t + 1|k}^{(s)}{f_{c,t + 1|t}}(x_{t + 1}^{(q)}|x_t^{(s)}){h_{r_{t + 1,c}^{(q)}r_{t,c}^{(s)}}}}}{{\mu _{c,t + 1|t}^{(q)}}}}  + 1 - {P_s}(x_t^{(s)})} \right],s = 1,...,{L_t}(c)
\end{equation}
where
\begin{equation}
\begin{array}{l}
\mu _{c,t + 1|t}^{(q)} = {\gamma _{c,t + 1}}(x_{t + 1}^{(q)}) + \sum\limits_{u = 1}^{{L_t}(c)} {w_{c,t|t}^{(u)}{h_{r_{t + 1,c}^{(q)}r_{t,c}^{(u)}}}\{ {P_s}(x_t^{(u)}){f_{c,t + 1|t}}(x_{t + 1}^{(q)}|x_t^{(u)})\} } \\
 + \sum\limits_{c''} {\sum\limits_{u = 1}^{{L_t}(c'')} {w_{c'',t|t}^{(u)}{\beta _{c,t + 1|t}}(x_{t + 1}^{(q)},r_{t + 1,c}^{(q)},c|x_t^{(u)},r_{t,c''}^{(u)},c'')} } 
\end{array}\end{equation}

The output of this step is a new set of particles 
$\{ w_{k - L|k - L}^{(s)},x_{k - L|k - L}^{(s)},r_{k - L|k - L}^{(s)},c_{k - L|k - L}^{(s)}\} _{s = 1}^{{L_{k - L}}}$. If we neglect the spawn targets, it is easy to see the PHD-JDTC smoothing is independent with each other across different class of targets.

\textbf{Step 5}:  Resampling after smoothing. To avoid the depletion of particles, we  resample the particles within each class again. Finally, the "smoothed" particles belongs to class $c$ of time $k-L$ becomes
$\{ \bar w_{k - L,c}^s,x_{k - L,c}^s,r_{k - L,c}^s\} _{s = 1}^{N_{k - L}^c}$

\textbf{Step 6}: State extraction and ordinal estimation. 
Then, the intra-class particle clustering is performed on different kinds of particles, and the motion state of the multi-target after smoothing is extracted, and all the particle weights are summed together, and the estimated number of target after smoothing is achieved.

Since the smoothing process involves a large number of interactions between particles, the computational complexity becomes higher. In order to achieve real-time requirements, the kd-tree method can be used here. K-d tree method considers the case where a large number of particles are densely distributed in space, and the "group-to-group" method can greatly increase the computation speed within tolerated error.

\section{Evaluation}~\label{sec:evaluation}

In order to verify the performance of the PHD-JDTC forward-backward smoothing proposed, we compare it with PHD-JDTC filter in target crossover and parallel motion scenarios. 

The attribute information of the target in~\cite{wei-a} is extracted from the signal-to-noise ratio of the radar measurement signal, which is assumed to be known. In real practice, however, the target true average signal-to-noise ratio is usually unknown or it varies within a certain interval. In~\cite{clark-a}, the likelihood function of the unknown SNR target is obtained by integrating the interval in the possible signal-to-noise ratio interval of the target. In this paper, we first divide the possible signal-to-noise ratio (SNR) of different categories of targets, and then use the likelihood function of the unknown SNR target as the clutter and the attribute likelihood function of each class target.

Suppose there are two different categories of targets in the surveillance area. The average SNR ratio of the targets is $45 dB$ and $25 dB$. We first divide the SNR interval into a high SNR interval $[30dB, 50dB]$ and a low SNR interval $[10dB, 30dB]$. Using the Rayleigh distribution amplitude model, the intensity of the signal is obtained using an envelope detector. The signal amplitude probability density of clutter and different categories of targets are 
\begin{equation}
{g_0}(a) = a\exp \left( {\frac{{ - {a^2}}}{2}} \right),a \ge 0
\end{equation}
\begin{equation}
{g_a}(a|{d_1},{d_2}) = \frac{{2\left( {\exp \left( {\frac{{ - {a^2}}}{{2\left( {1 + {d_2}} \right)}}} \right) - \exp \left( {\frac{{ - {a^2}}}{{2\left( {1 + {d_1}} \right)}}} \right)} \right)}}{{a\left( {\ln (1 + {d_2}) + \ln (1 + {d_1})} \right)}},a \ge 0
\end{equation}

 $a$ is the target signal amplitude, $[d_1, d_2]$ defines the possible average signal-to-noise ratio interval of a certain category of targets. The false alarm probability is 
\begin{equation}
p_{FA}^\tau  = \int_\tau ^{ + \infty } {{g_0}(a)da}
\end{equation} ,  the target detection probability is 
\begin{equation}
p_D^\tau (a|{d_1},{d_2}) = \int_\tau ^{ + \infty } {{g_a}(a|{d_1},{d_2})da} 
\end{equation}
 and $\tau$ is signal detection threshold.
 
The normalized probability density function over the signal detection threshold $\tau$ can be normalized, and the clutter and  attribute likelihood functions of the different classes of targets become:
\begin{equation}
g_0^\tau (a) = {{{g_0}(a)} \mathord{\left/
		{\vphantom {{{g_0}(a)} {p_{FA}^\tau }}} \right.
		\kern-\nulldelimiterspace} {p_{FA}^\tau }},a \ge \tau ,
\end{equation}

\begin{equation}
g_a^\tau (a|{d_1},{d_2}) = {{{g_a}(a|{d_1},{d_2}){\rm{ }}} \mathord{\left/
 {\vphantom {{{g_a}(a|{d_1},{d_2}){\rm{ }}} {p_D^\tau (|{d_1},{d_2})}}} \right.
 \kern-\nulldelimiterspace} {p_D^\tau (|{d_1},{d_2})}},a \ge \tau 
\end{equation}

\begin{figure}[!t]
	\centering
	\includegraphics[width=3.5in]{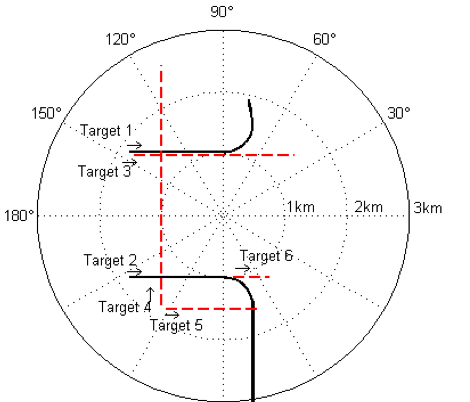}
	\caption{Simulation scenario of forward-backward PHD-JDTC.(---: type1, - -:type2) }
	\label{PFPHD_Trajectory}
\end{figure}

Assume a 2-D scenario with two synchronous sensors, which are located in $[1,0]km$ and $[0,0]km$, respectively. The surveillance area is $[0, 2\pi]rad \times [0, 15]km$. There are totally four targets which could be classified into two types. Target one (type 1) moves southeastwards with the initial position $[-10,10]km$ at time $k=0s$. At the same time, target two(type 2) and target three (type 1) moves northeastwards in parallel starting from $[-10,-10]km$ and $[-10,-10.5]km$. Target four (type 2) appears in $[1.4,8]km$ at time $120s$ and moves southwards. Target four disappeared at time $480s$ while the rest targets disappeared at $360s$. The trajectories of the targets are shown in Fig.~\ref{PFPHD_Trajectory}. 

The target state takes the form of ${X_k} = [x_k,\dot{x}_k,y_k,\dot{ y}_k,\omega_k ]^T $. The time evolution of the state $X_k$ of target movement is given by ${X_k} = {F_k}{X_{k - 1}} + {v_k}$. The movement models of the target is described by two models: constant-velocity (CV) model and coordinated turn (CT) model with turn rate unknown. For CV model:
\begin{equation}
{F_{k,CV}} = diag\!\left(\! {\left[ {\begin{array}{*{20}{c}}
1&{dT}\\
0&1
\end{array}} \right],\left[ {\begin{array}{*{20}{c}}
1&{dT}\\
0&1
\end{array}} \right],0}\! \right) \,,
\end{equation}
and ${v_{k,CV}}\sim N(\cdot;0,Q_{k,CV})$, and $dT=6s$ is the sample time step.
\begin{equation}
{Q_{k,CV}} = diag\left( {\left[ {\begin{array}{*{20}{c}}
{\frac{{d{T^3}l}}{3}}&{\frac{{d{T^2}l}}{2}}\\
{\frac{{d{T^2}l}}{2}}&{dTl}
\end{array}} \right],\left[ {\begin{array}{*{20}{c}}
{\frac{{d{T^3}l}}{3}}&{\frac{{d{T^2}l}}{2}}\\
{\frac{{d{T^2}l}}{2}}&{dTl}
\end{array}} \right],0} \right) \,,
\end{equation}
where $l = 1{m^2}/{s^3}$.

For the CT model with turning rate unknown:
\begin{equation}
{F_{k,CT}} =\!\left[\! {\begin{array}{*{20}{c}}
1&{\frac{{\sin ({w_{k - 1}}dT)}}{{{w_{k - 1}}}}}&0&{ - \frac{{\cos ({w_{k - 1}}dT)}}{{{w_{k - 1}}}}}&0\\
0&{\cos ({w_{k - 1}}dT)}&0&{ - \sin ({w_{k - 1}}dT)}&0\\
0&{ - \frac{{\cos ({w_{k - 1}}dT)}}{{{w_{k - 1}}}}}&1&{\frac{{\sin ({w_{k - 1}}dT)}}{{{w_{k - 1}}}}}&0\\
0&{\sin ({w_{k - 1}}dT)}&0&{\cos ({w_{k - 1}}dT)}&0\\
0&0&0&0&1
\end{array}} \!\right] \,,
\end{equation}
and ${v_{k,turn}}\sim N(\cdot;0,{Q_{k,CT}})$.
\begin{equation}
{Q_{k,CT}} = diag\!\left(\! {\left[\!\! {\begin{array}{*{20}{c}}
{\frac{{d{T^3}{l_1}}}{3}}&{\frac{{d{T^2}{l_1}}}{2}}\\
{\frac{{d{T^2}{l_1}}}{2}}&{dT{l_1}}
\end{array}} \!\!\right],\left[\!\! {\begin{array}{*{20}{c}}
{\frac{{d{T^3}{l_1}}}{3}}&{\frac{{d{T^2}{l_1}}}{2}}\\
{\frac{{d{T^2}{l_1}}}{2}}&{dT{l_1}}
\end{array}} \!\!\right],dT{l_2}}\! \right) \,,
\end{equation}
where ${l_1} = 1{m^2}/{s^3}$ and ${l_2} = 0.1ra{d^2}/{s^3}$.
The transition matrix between two motion models is $\pi  =\! \left[\!\! {\begin{array}{*{20}{c}}
{0.8}&{0.2}\\
{0.2}&{0.8}
\end{array}} \!\!\right]$.

\begin{figure}[!t]
	\centering
	\includegraphics[width=3.5in]{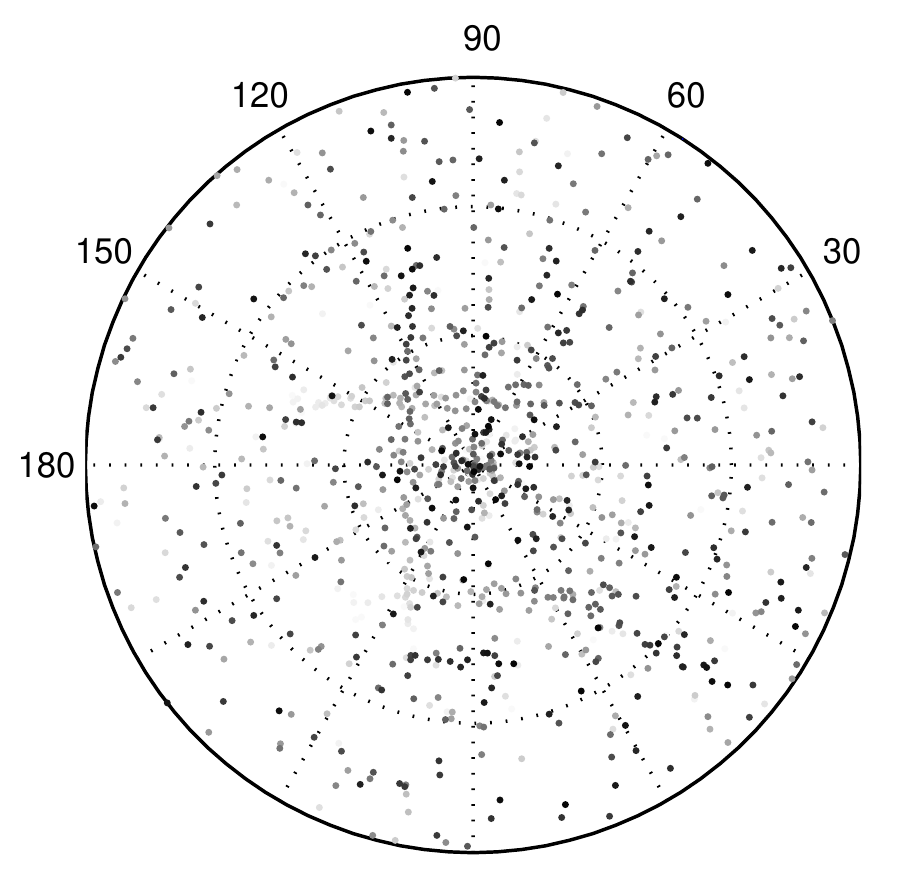}
	\caption{Measurements of sensors with clutter}
	\label{measurments}
\end{figure}

Assuming that the observation equation for $i$th sensor in polar coordinates is
\begin{equation}
z_{k,i}^{} = \left[ {\begin{array}{*{20}{c}}
   {\sqrt {{{\left( {{x_{k,i}} - {p_{x,i}}} \right)}^2} + {{\left( {{y_{k,i}} - {p_{y,i}}} \right)}^2}} }  \\
   {{\text{atan}}\left( {\frac{{{y_{k,i}} - {p_{y,i}}}}
{{{x_{k,i}} - {p_{x,i}}}}} \right)}  \\
 \end{array} } \right] + w_{k,i}
\end{equation}
The measurements observed by the sensors is shown in Fig.~\ref{measurments}. 
 the measurement noise follows Gaussian distribution $w_k^i\sim N(\cdot;0,R_w^i)$ with the covariance of measurement from two sensors $R_w^1 = R_w^2 = diag([{(300m)^2},{\left( {{\pi  \mathord{\left/
 {\vphantom {\pi  {180}}} \right.
 \kern-\nulldelimiterspace} {180}}rad} \right)^2}])$.

At time $k$, the intensity of birth target ${\gamma _{k,1}}({x_k})$  for type 1 target is
\begin{equation}
\begin{array}{l}
{\gamma _{k,1}}({x_k}) = 0.01N\left( {{x_k};{{\left[ {{\rm{ -1500,0,1000,0,0}}} \right]}^T},diag\left( {\left[ {{\rm{1,100,1,100,1}}{{\rm{0}}^{ - 6}}} \right]} \right)} \right)\\
 + 0.01N\left( {{x_k};{{\left[ {{\rm{ -1500, 0, -1000, 0, 0}}} \right]}^T},diag\left( {\left[ {{\rm{1, 100, 1, 100, 1}}{{\rm{0}}^{ - 6}}} \right]} \right)} \right)
\end{array}
\end{equation}

The intensity of birth target ${\gamma _{k,2}}({x_k})$  for type 1 target can be described by

\begin{equation}
\begin{array}{l}
{\gamma _{k,2}}({x_k}) = 0.01N\left( {{x_k};{{\left[ {{\rm{ -1500, 0, 1000, 0, 0}}} \right]}^T},diag\left( {\left[ {{\rm{1, 100, 1, 100, 1}}{{\rm{0}}^{ -7}}} \right]} \right)} \right)\\
 + 0.02N\left( {{x_k};{{\left[ {{\rm{ -1000, 0,  -1500, 0, 0}}} \right]}^T},diag\left( {\left[ {{\rm{1, 100, 1, 100, 1}}{{\rm{0}}^{ - 7}}} \right]} \right)} \right)
\end{array}
\end{equation}
The intensity of spawn target is 

\begin{equation}{\beta _{k|k - 1}}({x_k},{r_k},{c_k}|{x_{k - 1}},{r_{k - 1}},{c_{k - 1}}) = {\lambda _\beta }{f_\beta }({x_k}|{x_{k - 1}}){f_{k|k - 1}}({r_{k,{c_k}}}|{r_{k - 1,{c_{k - 1}}}})b({c_k}|{c_{k - 1}})\end{equation}

where 
${\lambda _\beta } = 0.01$,
\begin{equation}{f_\beta }({x_k}|{x_{k - 1}}) = N\left( {{x_k}|{x_{k - 1}},diag\left( {\left[ {{\rm{100, 1, 100, 1, 1}}{{\rm{0}}^{ - 8}}} \right]} \right)} \right)\end{equation}

\begin{equation}{f_{k|k - 1}}({r_{k,{c_k}}}|{r_{k - 1,{c_{k - 1}}}}) = \left[ {\begin{array}{*{20}{c}}
{0.9}&{0.1}\\
{0.1}&{0.9}
\end{array}} \right]\end{equation}
$b({c_k}|{c_{k - 1}}) = \left[ {\begin{array}{*{20}{c}}
{0.5}&{0.5}\\
{0.5}&{0.5}
\end{array}} \right]$
The probability of target survival is ${P_s} = 0.99$. The sequential Monte Carlo approximation implementation have 500 particles for each existing target, while 500 particles are used to initialize each new-born target.

\begin{figure}[!t]
	\centering
	\includegraphics[width=3.5in]{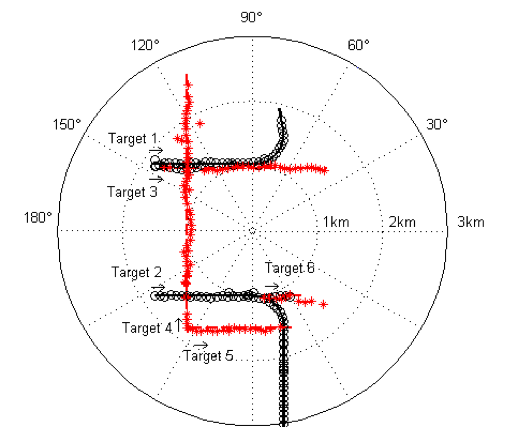}
	\caption{Estimated trajectories with forward-backward PHD-JDTC smoothing}
	\label{fig:filterResult}
\end{figure}

\subsection{Simulation Results}
Fig.~\ref{fig:filterResult} shows the estimated trajectories and classes with forward-backward PHD-JDTC smoothing. The proposed algorithm is able to filter out the state of targets, as well as their class.

\begin{figure}%
	\centering
	\subfloat[][Target number error comparison for type 1 targets]{\includegraphics[width=3in]{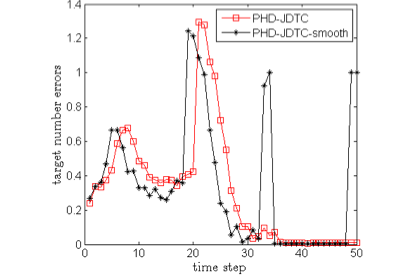}\label{fig:firstclass}} 
	\subfloat[][Target number error comparison for type 2 targets]{\includegraphics[width=3in]{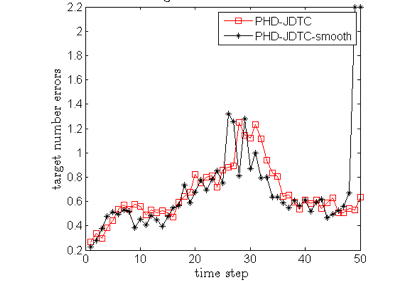}\label{fig:secondclass}}
	\caption{Target number error for all type of classes}%
	\label{fig:targetnumber}%
\end{figure}

\begin{figure}%
	\centering
	\subfloat[][OSPA of type 1 targets]{\includegraphics[width=2.5in]{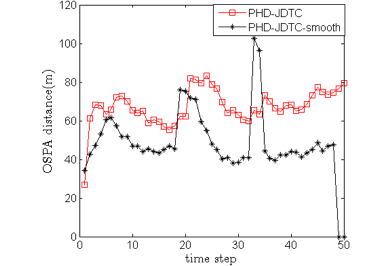}\label{fig:firstclassOSPA}} 
	\subfloat[][OSPA of type 2 targets]{\includegraphics[width=2.5in]{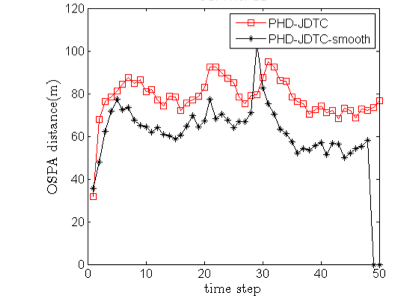}\label{fig:secondclassOSPA}}
    \subfloat[][OSPA of all targets]{\includegraphics[width=2.5in]{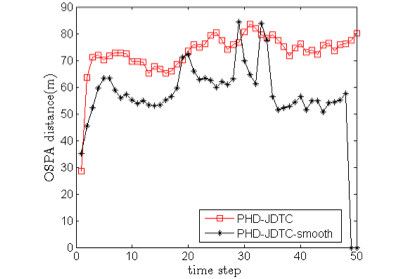}	\label{fig:allclassOSPA}}
	\caption{OSPA distance for all type of classes}%
	\label{fig:OSPA}%
\end{figure}

The OSPA distance for different approaches are provided in Fig.~\ref{fig:OSPA}. We find that the algorithm's estimated OSPA distance for different categories of targets is reduced by up to 40\% compared with PHD-JDTC. Therefore, the PHD-JDTC forward-backward smoothing proposed outperforms the state-of-art PHD-JDTC filter in tracking accuracy. 
After smooth, the OSPA distance for target with high SNR ratio is smaller compared with that of low SNR ratio. The reason is that the target with high SNR ratio is more distiguish from clutter and less affected by noise measurements. The target with low signal-to-noise ratio is close to the clutter SNR ratio interval, so the capability to distinguish between target and noise is poor.

The estimation of target number in different categories after smooth is basically the same as that of after smoothing. This is consistent with the theoretical analysis in ~\cite{mahler-a}, smoothing cannot improve the cardinality estimation accuracy. From Fig.~\ref{fig:targetnumber}, we can see that smoothing has hysteresis properties. The disappearance of target occurs beforehand, causing the target number to be inaccurate during target disappears.

Our result also shows that the introduction of category information can effectively deal with the joint target detection, tracking and classification even target has crossover or parallel motions. It is noted that the time complexity of our approach is smaller than the regular PHD smoothing, which does not consider class information.

\section{Conclusion and Future Work}\label{sec:conclusion}

In this paper, we derived the generalized PHD smoothing characterized by extending the general mode to PHD filtering and smoothing. Through a “top-down” method, the general mode augmented forward-backward PHD filter is derived. The forward-backward PHD-JDTC filter has explicit structure and its SMC implementation is presented. The evaluation results show the improved performance of our approach in target detection, tracking and classification compared with the state-of-art PHD-JDTC filter. The simulation scenario we  considered is a two dimension case. Whereas, it is quite easy to be extended to 3-D by taking into account the pitch angle measuremented by sensors. Furthermore, It is well known that the PHD filter is not able to output the track information, due to the lack of data association. Therefore, it is recommended to incorporate the track label based on our approach to have the track information maintained~\cite{lin-a}. Moreover, Gaussian Mixture PHD is the closed form implementation of PHD filter under the assumption of linear and Gaussian noise. The corresponding Gaussian Mixture PHD-JDTC smoothing is worth further study.

\bibliographystyle{IEEEtran}
\bibliography{ref}

\end{document}